\documentstyle[psfig,astron]{lamuphys}

\def\spose#1{\hbox to 0pt{#1\hss}}
\def\lta{\mathrel{\spose{\lower 3pt\hbox{$\mathchar"218$}}
     \raise 2.0pt\hbox{$\mathchar"13C$}}}
\def\gta{\mathrel{\spose{\lower 3pt\hbox{$\mathchar"218$}}
     \raise 2.0pt\hbox{$\mathchar"13E$}}}
\def\HI{H\,{\small I}}

\begin{document}

\title{High redshift radio galaxies with the VLT} 

\author{Huub R\"ottgering \&  George Miley} 

\institute{Leiden Observatory,\\ P.O. Box 9513, \\ 2300 RA Leiden, \\
The Netherlands}
\maketitle
  
\section{Introduction} 

The study of high-$z$ radio galaxies (HZRGs) has matured considerably during
the last decade.  Large numbers of $z>2$ radio galaxies have been 
found  and have been studied in considerable detail.

During the mid-eighties there were indications from statistical
analyses of radio source counts that there might be a cutoff in the
space density of powerful radio sources above a redshift of $z\sim
2$ (e.g. Windhorst 1984). \nocite{win84a} However, it has become
clear that a substantial population of radio galaxies with $z>2$ does
exist. The first radio galaxy discovered with a measured redshift
beyond 2 was 4C40.36 at $z=2.269$ (Chambers et al. 1988).
\nocite{cha88a} This was soon followed by two radio galaxies 
at even more extreme distances (0902+34
at $z=3.4$, Lilly 1989 \nocite{lil89} and 4C41.17 at $z=3.8$, Chambers
et al. 1990). \nocite{cha90c}  At the
moment the record holder is 6C0140+326 at $z=4.41$ (Lacy et al. 1996).
\nocite{lac96a}

\begin{figure} 
\centerline{
\psfig{figure=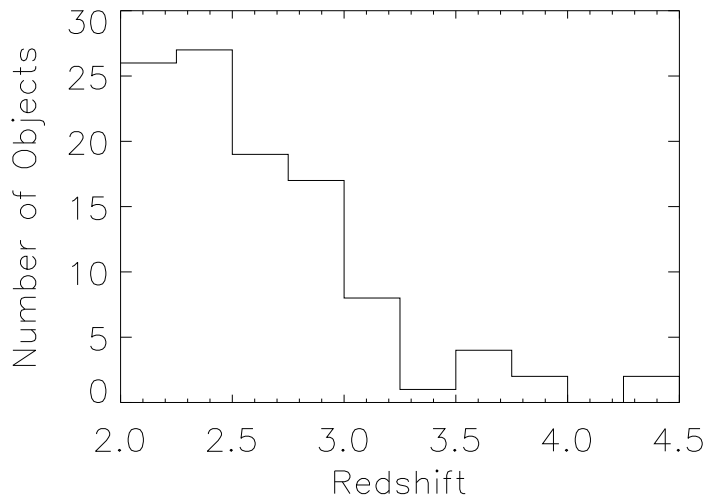,width=11cm}}
\caption{\label{zdistr}
Redshift distribution of the $z>2$ radio galaxies known to the 
authors in mid-1996.}
\end{figure}

Not only has the maximum distance out to which radio galaxies are 
observed been extended  dramatically, but also the number of known
galaxies with $z>2$ has grown substantially to 
well above 100 (e.g. Spinrad 1995, Eales and Rawlings 1996, McCarthy et
al. 1996, R\"ottgering et al. 1996a).
\nocite{spi95b,eal96,mcc96,rot96e} In Fig. \ref{zdistr} the redshift
distribution of 106 $z>2$ galaxies that are known to the authors
(mid-1996) is shown. This histogram is of that 
given by Spinrad (1995) showing the 79 $z>2$ radio galaxies that were known
to him in mid-1994.  The number of known distant radio galaxies has
increased by a third over this two year period. 

In this review we will address the importance of the VLT for studies
of samples of distant radio galaxies.  We shall first briefly discuss
the importance of radio galaxies for cosmological studies. Then we
shall summarize  the various emitting components associated with these
objects and  comment on the possibilities offered by the VLT
for detailed observations of these components and the 
radio/optical alignment effect.
Next, the prospects for  detecting and studying galaxy
clusters around high redshift radio galaxies are  briefly
discussed. The case of 1138$-$262 ($z=2.15$) is discussed as an example
of a possible proto-galaxy that might be situated in a cluster-type of
environment.  Finally, we outline some of the steps that should be taken
to optimize use of the VLT in this area.

For the most complete review of distant radio galaxies,  as
per 1993, we refer the reader to McCarthy (1993).  \nocite{mcc93}

\section{Importance}

As a probe of the distant universe, distant radio galaxies offer a
wide range of distinct components that can be
studied. These include stars, AGN, hot (X-ray emitting)
gas, ionized gas, neutral (HI) gas, molecular gas, dust and
relativistic plasma. Some components have merely been detected (dust),
while others have been extensively studied (ionized gas, neutral (HI)
gas, relativistic plasma). 
It is commonly accepted that these objects
contain stars and AGN, but their existence have been inferred by
indirect methods. Other components have not yet been detected
in the highest redshift objects (molecular
gas and hot (X-ray emitting) gas), 
but it
is clear that only modest advances in the performance of
instrumentation should allow detection of these components.

Radio galaxies have advantages over quasars as cosmological 
probes in  that at least
four of these emitting components can be spatially resolved by ground based
telescopes.  Furthermore, due to the absence of a bright nucleus, the
study of components other than the nucleus itself is less challenging.

One of the aims of studying these distant sources is to better
understand processes related to galaxy formation. Possibly the best
hint that radio source activity in these distant objects is related to
galaxy formation is the strong evolution of the radio luminosity
function.  The number density 
of the most  powerful steep spectrum radio sources
(e.g. P$_{2.7} = 10^{27}$ W Hz$^{-1}$ sr$^{-1}$) increases
by  a factor 1000 from $z\sim0$ up to $z\sim 2-3$ (Dunlop and Peacock
1993).  \nocite{dun93} Such strong evolution is a 
natural consequence in hierarchical models for structure formation 
(Haehnelt and Rees 1993). \nocite{hae93}

Finally, radio galaxies are powerful tracers of galaxy clustering at
high redshift.  There are a number of observational hints that the
highest-$z$ radio galaxies are located in relatively dense
environments.  We shall  discuss this in some detail in section 5.

\section{Components} 

In Table \ref{overview} we provide an overview of the mass associated
with some fo the components of HZRGs.   Most of these estimates are
highly model-depended and should therefore treated with caution.

\begin{table}
\begin{center} 
\begin{tabular}{lcl} 
\hline \hline
Components & Total Mass & \qquad \qquad Reference  \\ 
           &  M$_\odot$ &           \\
\hline \\ 
Stars         &$10^{12}$ &4C41.17, $z=3.8$, Chambers et al. (1990) \nocite{cha90c}\\
Ionized Gas   &$10^9$    &1243+036, $z=3.6$, van Ojik et al. (1996a) \nocite{oji96a}\\
Neutral (HI) gas 
              &$2 \times 10^7$    &0943$-$242, $z=2.9$,  R\"ottgering et al. (1995) \nocite{rot95a} \\
Molecular gas &$<10^{11}$&Evans et al. (1996); van Ojik et al. (1996b)  \nocite{eva96,oji96b}   \\
Dust          &$3 \times 10^8$    &4C41.17, $z=3.8$, Dunlop et al. (1994) \nocite{dun94} \\
Hot (X-ray emitting) gas 
              &$2 \times 10^{12}$   &3C356, $=1.079$,  Crawford and Fabian (1993) \nocite{cra93}\\ 
%PRINT,0.01* 4 * (257/2. * 3D21)^3 *1.7D-24/2D33
\hline 
\end{tabular} 
\end{center} 
\caption{\label{overview} Estimated masses
for some of the components associated with distant radio galaxies. 
A Hubble constant of $H_0=50$ km s$^{-1}$ Mpc$^{-1}$ and 
a density parameter of $\Omega = 1 $ is used.}  
\end{table} 

Below we will discuss characteristics of the components that 
can be studied with the VLT and give some 
recent results  indicating  what progress 
can be made using the VLT.

\subsection {Stars} 

It is clear that characterization of the stellar population in distant
galaxies is a prime task for the next generation of optical
telescopes.  However, the first issue to be addressed is whether there is
observational evidence that these systems do indeed contain stars.  At
low-$z$, powerful radio sources are located in massive ellipticals. By
analogy, one would expect this also to be the case at high redshifts. However,
optical imaging studies of the morphology of $z>2$ galaxies showed
that a large fraction of these objects are   clumpy and have   its  main
optical axis aligned with the axis of the radio emission
(e.g. McCarthy 1993).  This is unlike nearby ellipticals and therefore,
before the stellar content of these systems can be properly modelled,
it has to be established whether these systems do contain stars and, if
so, what fraction of the UV/optical/infrared continuum emission is due
to stars. We will return to this issue in section \ref{sec-inter}.  Here
we will concentrate on observational evidence that these systems do
contain stars.

Until recently the best evidence that these systems do contain a
massive population of stars came from infrared imaging.  The alignment
effect and clumpiness is much less prominent at these relatively long
wavelengths and the Hubble K-diagram shows a continuity from low to
high-$z$ and a relatively low dispersion
(e.g. McCarthy 1993). All this can be readily
explained as related to the existence of a massive population of
relatively old stars (for alternative views see Eales and Rawlings
1996). \nocite{eal96} 

The best direct evidence for stars comes from
recent KECK spectroscopy of a radio galaxy at $z=3.54$ (Dey et al. in
prep.).  In addition to the usual emission lines (Ly$\alpha$, C IV and
He II) it showed very strong UV absorption lines.
This absorption spectrum very much resembles that of the two nearby
Wolf-Rayet galaxies NGC 1741 (Conti et al. 1996) \nocite{con96} and NGC
4214 (Leitherer et al. 1996) \nocite{lei96} as well as those of the
starforming galaxies at $3 < z < 3.5 $ that Steidel et al. (1996)
\nocite{ste96} have recently discovered.  This similarity clearly
indicates that at least one distant radio galaxy contains a massive
population of young stars that  have  been produced during a
recent burst of star formation.

The VLT with its proposed optical and infrared instrumentation should
allow detailed studies of spectral energy distribution of these
systems, including stellar absorption lines.  With this information it
should be possible to separate the contributions of  stellar and
non-stellar light to the morphology of the galaxy.  This will allow
important parameters of these systems to be determined, including
stellar content, metallicity, SFR and SF history.  The high resolution
capabilities of the VLT should allow a number of these parameters to
be determined as a function of location within the galaxy. With this
information it should be possible to study the distribution of SF over the
galaxy and observationally establish the epoch of formation of these
galaxies.

\subsection {AGN}  

Since distant radio galaxies all emit powerful radio emission, it is
likely that these objects contain, either now or in the past, active
nuclei.  During this meeting, Cimatti discussed how observations of
scattered light will not only contribute to our knowledge of the
scattering medium, but also to that of the nature of the AGN.

It is likely that dust absorbs broad UV emission lines associated
with the supposed nucleus in the very centre of the HZRGs.  Since dust
obscuration is less severe in the infrared, searches for broad
emission lines and compact central objects should be carried out in 
this waveband.  In addition, the high resolution imaging of
radio loud quasars should give detailed information on the galaxies
hosting the quasars. A detailed comparison between these host and
those of radio galaxies should provide important further constrains on
models that attempt to unify quasars and radio galaxies.

\subsection {Ionized gas} 

The Ly$\alpha$ emission of distant  radio galaxies is often
spectacular (e.g. McCarthy 1993).  \nocite{mcc93} It can be as
luminous as $10^{44}$ erg s$^{-1}$ and extend up to 100 kpc ($ \sim
10''$ at $z=2.5$).  These halos could well trace the reservoir of gas
from which the galaxies are forming.

Studies of the velocity structure of the gas show that there are two
distinct components.  The inner region has a relatively high velocity
dispersion ($\sim 1000$ km s$^{-1}$) and there is clear evidence that 
this 
is induced by hydrodynamical interactions of the
radio jet with the gas. In some sources such interaction is directly
observed; in 1243+036 ($z=3.6$) the Ly$\alpha$ 
gas is displaced both spatially and in
velocity at the location of a strong bend in the radio jet
(van Ojik et al. 1996a). \nocite{oji96a}  Further
evidence for such strong interaction comes from the correlated
distortions  of the radio and emission line morphologies
seen in samples of HZRGs (van Ojik et al. 1996c).  \nocite{oji96b} The second
gas component is the emission line gas outside the main radio source structure,
which  has a much lower velocity dispersion ($\sim 250$ km s$^{-1}$).  In 
the case of 1243+036 we have direct evidence that the gas shows
ordered  motion, possible due to rotation of a protogalactic gas disk at
$z = 3.6$ out of which the galaxy associated with 1243+036 is forming.
A gravitational origin of the rotation of such a large disk implies a
mass of $\sim 10^{12} \sin ^{-2}(i)$ M$_{\odot}$, where $i$ is the
inclination angle of the disk with respect to the plane of the sky.

The distinction between the inner and outer halos has been observed
using 4-m telescopes with integration times of order 1 night, about
the maximum realistically feasible.  An interesting challenge for an
8-m telescope would be to determine the maximum extent of the Ly$\alpha$
halos. The largest sizes measured so far are up to 150
kpc. Is this a true cutoff in the distribution or is it  possible to
trace the Ly$\alpha$ halo out to distances of half a megaparsec?

A very important question concerns the ionisation of the emission line gas.
For a number of reasons it is likely that ionisation by a beam from 
a hidden quasar is playing a role: (i) some HZRGs have a 
Ly$\alpha$ emitting region that has a cone shaped
morphology reminiscent of such a scenario, (ii) the integrated line
ratios are well reproduced by nuclear photoionisation models and (iii)
there is a tendency for   the emission line region to be aligned with
the main axis of the radio source (e.g. the optical beam).  However, 
there are a number of problems with this simple picture, the most
important being that the outer contours of the Ly$\alpha$
halos are elliptical in shape, and that therefore a significant
fraction of halo emission comes from regions perpendicular to the
radio axis, ie. not illuminated by the supposed beam.  An important
constraint on the ionisation mechanisms would be to measure line ratios
in these outer  regions.
 
With detailed studies of the morphology, dynamics and ionisation
mechanisms of the halos gas, it should be possible to test scenarios
for the origin of the gas.  It has been suggested that the gas was
expelled during a major starburst that the galaxy underwent during its
formation.  Alternatively, the gas might be indicative of a massive
cooling flow that provides a significant fraction of the material from
which the galaxy is forming.

\subsection {Neutral gas}  

There are a number of ways to search for and subsequently study
neutral gas associated with distant galaxies.  One method is to
measure the redshifted 21 cm absorption line against the radio
continuum.  The only distant radio galaxy for which this has been
done  is 0902+34 ($z=3.4$). Uson et al. (1991) \nocite{uso91} 
found an absorber in
this system with a column density of $4.4 \times 10^{22}$ atoms
cm$^{-2}$, assuming a spin temperature of $10^4$ K. \nocite{uso91} The
existence of this absorber was confirmed by Briggs et
al. (1993)  \nocite{bri93} and de Bruyn et al. (1995).  \nocite{bru95}
It is likely that more cases will be discovered by the new tunable
radio receivers of the Westerbork Radio Telescope (see {\tt
http://www.nfra.nl/nfra/wsrt\_info.html}).

A second method is studying the deep narrow troughs that often
``disfigure'' the Ly$\alpha$ profiles.  High--resolution spectra show
that, in some cases, these features are too sharp to be explained as
separate kinematic components of the emission, but that they are
definitely due to absorption by neutral hydrogen along the line of sight.
We have analysed deep high resolution spectra for a sample of 18
distant radio galaxies (van Ojik 1995; van Ojik et al. 
1996c) \nocite{oji95d,oji95b} and \HI\ absorption features appear
widespread in the Ly$\alpha$ profiles.  11 radio galaxies out of the
sample of 18 have strong ($>10^{18}$ cm$^{-2}$) \HI\ absorption.
Since, in most cases, the Ly$\alpha$ emission is absorbed over the
entire spatial extent (up to 50 kpc), the absorbers must have a
covering fraction close to unity. Given the column densities and
spatial scales of the absorbing clouds, the typical \HI\ mass of these
clouds is $\sim 10^8$ M$_{\odot}$.

\begin{figure} 
\centerline{
\psfig{figure=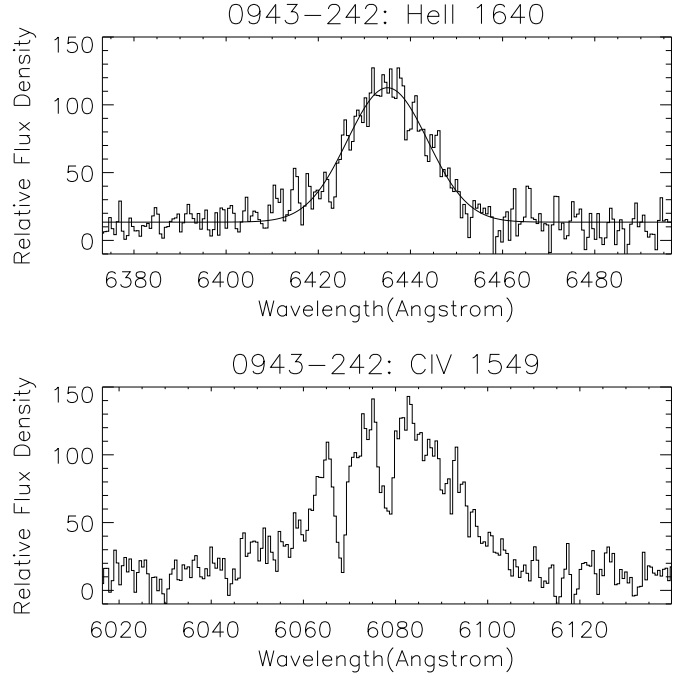,width=10cm}}
\caption{\label{0943-CIV} Parts of the high resolution AAT spectrum
(1.5 \AA) of the He II 1640 region (upper) and the CIV 1549 region
(lower) of the distant radio galaxy 0943$-$242 at $z=2.9$). The C IV
line shows absorption due to the CIV 1548/1551 doublet.} 
\end{figure}

On the source with one of the deepest and best defined HI absorption
systems (0943$-$242, $z=2.9$, see also Fig. 4), we have carried out
deep high resolution (1.5 \AA) spectroscopy on the C IV and He II line
using the AAT telescope.  In Fig.  \ref{0943-CIV} we show the
resulting spectra.  The He II line does not show absorption.  This is
expected since it is a non-resonant line. The C IV line shows
absorption due to the CIV 1548/1551 doublet.  We have fitted the line
profile with a combination of a Gaussian (for the emission) and two
(coupled) Voigt functions (for the absorption).  The column density
for the absorber is $10^{14.4}$ cm$^{-2}$.  Combined with the measured
column density for the HI absorber ($10^{19}$ cm$^2$), this indicates
that the spatially extended absorber is metal enriched. It further
shows that such absorption measurement is a good tool to study
extended slabs of neutral gas at high redshift.  Unfortunately,
0943$-$242 is one of the few objects for which this kind of work is
possible with 4-m class telescopes, since it is among the few that 
have  both strong HI absorption and strong CIV emission.

With the availability of the VLT there is the prospect of 
conducting detailed
studies of such systems.  Questions that should be addressed  concern the
dynamics and morphology of the neutral gas.  For example, could the neutral 
gas  be in  a rotating system or is it the product of 
colliding proto-galaxies? 
Such observations should also address the origin of the gas and
its ultimate fate as possible food for forming the stellar populations
of the galaxy.

\subsection{Dust}   

\begin{figure} 
\centerline{
\psfig{figure=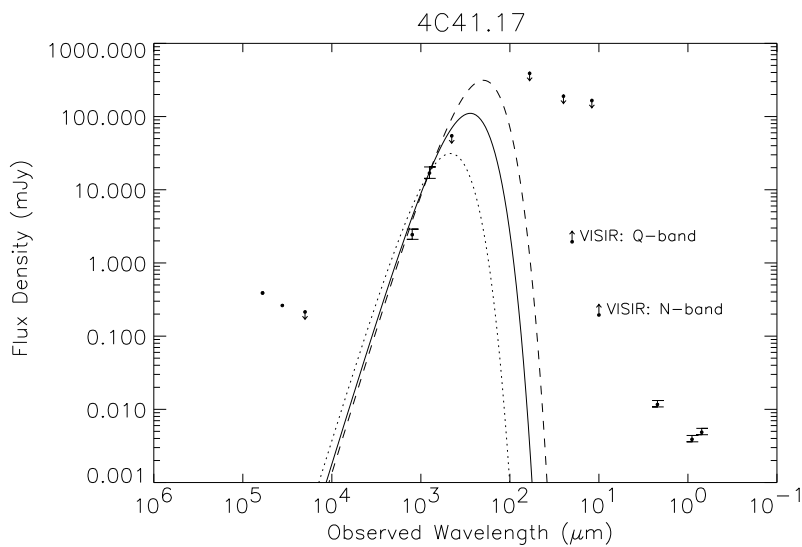,width=10.5cm}
}
\caption{\label{dust} Spectral energy distributions of the radio
galaxy 4C41.17 ($z=3.8$); the radio emission is for the
core only.  The dotted, solid, and dashed lines represent isothermal
grey-body emission with an emissivity index $\beta = 2$, for dust at
temperatures of 30\,K, 50\,K and 70\,K respectively. Data are taken
from the compilation by Hughes (1996).  Also  indicated is the 6
$\sigma$ RMS in an 8 hour observation for N-band ($8-13$ $\mu$ m) and
Q-band ($16-24$ $\mu$m) using VISIR,  the mid infrared instrument
on the VLT.} 
\end{figure}

The existence of dust associated with distant radio galaxies is well
established.  The most direct method is measuring the dust emission at
submillimetre wavelengths. In Fig. \ref{dust} we show the spectral
energy distribution of the radio galaxy 4C41.17 ($z=3.8$) (from a
compilation of Hughes 1996) \nocite{hug96} indicating that this object
contains $10^8$ M$_\odot$ of dust.  Other indirect measurements
confirm that HZRGs can indeed contain massive amounts of dust,
including (i) the optical/UV polarisation measurements, (ii) the
clumpy optical continuum morphologies as compared to those in the
infrared and (iii) the Ly$\alpha$/H$\alpha$ emission line ratios.

The amount of dust probably greatly varies from object to object. This
is exemplified by two objects (TX0211$-$122, van Ojik et al. 1994 and MG
1019+0535, Dey et al. 1995) \nocite{oji94a,dey95} out of an estimated 60
HZRGs that have Ly$\alpha$ very much fainter with respect to
the high ionisation lines than in typical high redshift radio
galaxies.  This suggests that these galaxies are undergoing a vigorous
starburst producing a copious amount of dust that  attenuates the Ly$\alpha$
emission. 

An interesting topic for the VLT will be to compare the spatial
distribution of H$\alpha$ to Ly$\alpha$, thereby estimating the
distribution of dust through the whole galaxy.  Detailed spectral
polarisation measurements with the VLT is another powerful tool for studying the
dust distribution. This is discussed by Cimatti in this proceedings.

Finally,  it it is interesting to consider whether the thermal infrared
instrument VISIR (VLT Imager and Spectrometer for mid InfraRed) 
that will be mounted on UT2 is
sensitive enough to detect dust at these high redshifts.  The 6
$\sigma$ RMS in an 8 hour observation is 0.2 mJy for N-band ($8-13$
$\mu$m) and 2 mJy for Q-band ($16-24$ $\mu$m).  We have plotted these
two limits in Fig. \ref{dust}.  From this it is clear that dust is only
detectable if there is a warm component (${\rm T}>100$ K) comparable in mass
to the colder dust.  IRAS has established that such warm dust
components are a common feature of nearby radio loud AGN (e.g. 3C390.3
Miley et al. 1984). \nocite{mil84b}   ISO should provide more information.
VISIR has the great advantage over ISO in studying such warm dust
because of its superb resolution.

\section{The alignment effect} 
\label{sec-inter} 
Several years ago it came as a big shock when it was discovered that,
unlike the case for nearby radio galaxies, the optical/IR continuum 
emission 
radio emission of $z>
0.6$ radio galaxies is roughly aligned with the  radio emission 
(Chambers, Miley and van Breugel 1987; McCarthy et al. 1987).
\nocite{mcc87a,cha87} During the last 10 years a large number of
explanations of this aligned emission have been proposed, the three
most promising being scattering of light from a hidden quasar by
electrons or dust (Tadhunter et al. 1989; Fabian 1989),
\nocite{tad87,fab89a} star formation stimulated by the radio jet as
it propagates outward from the nucleus (Chambers, Miley and van
Breugel 1987; McCarthy et al. 1987; De Young 1989; Rees 1989; Begelman
and Cioffi 1989) \nocite{you89,cha87,mcc87a,ree89a,beg89} and nebular
continuum emission from the emission line gas (Dickson et al.
1995). \nocite{dic95a}

\begin{figure}
\centerline{
\psfig{figure=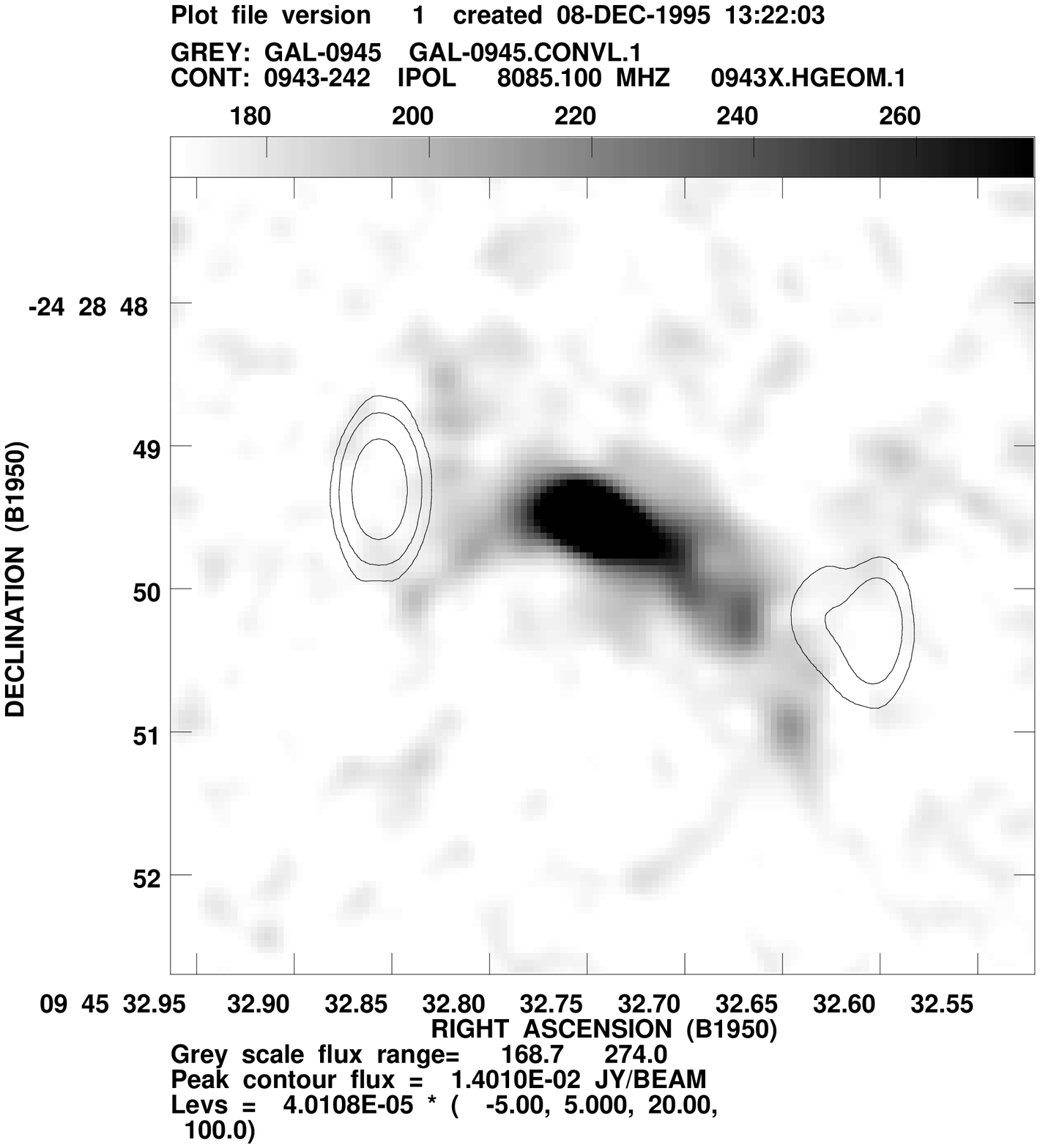,width=8cm,clip=}
}
\caption{\label{0943}
The grey-scale is the HST image of  0943$-$242 ($z=2.9$) 
through the  F702W filter 
with a total integration time of 5300 sec.
The contours show the VLA A-array total intensity radio map 
at 8.2 GHz  with a resolution of $0.25''$. 
The contours are at (0.2,0.8,4) mJy. 
}

\end{figure}

\begin{figure}
\vspace*{0.35cm} 
\centerline{\psfig{figure=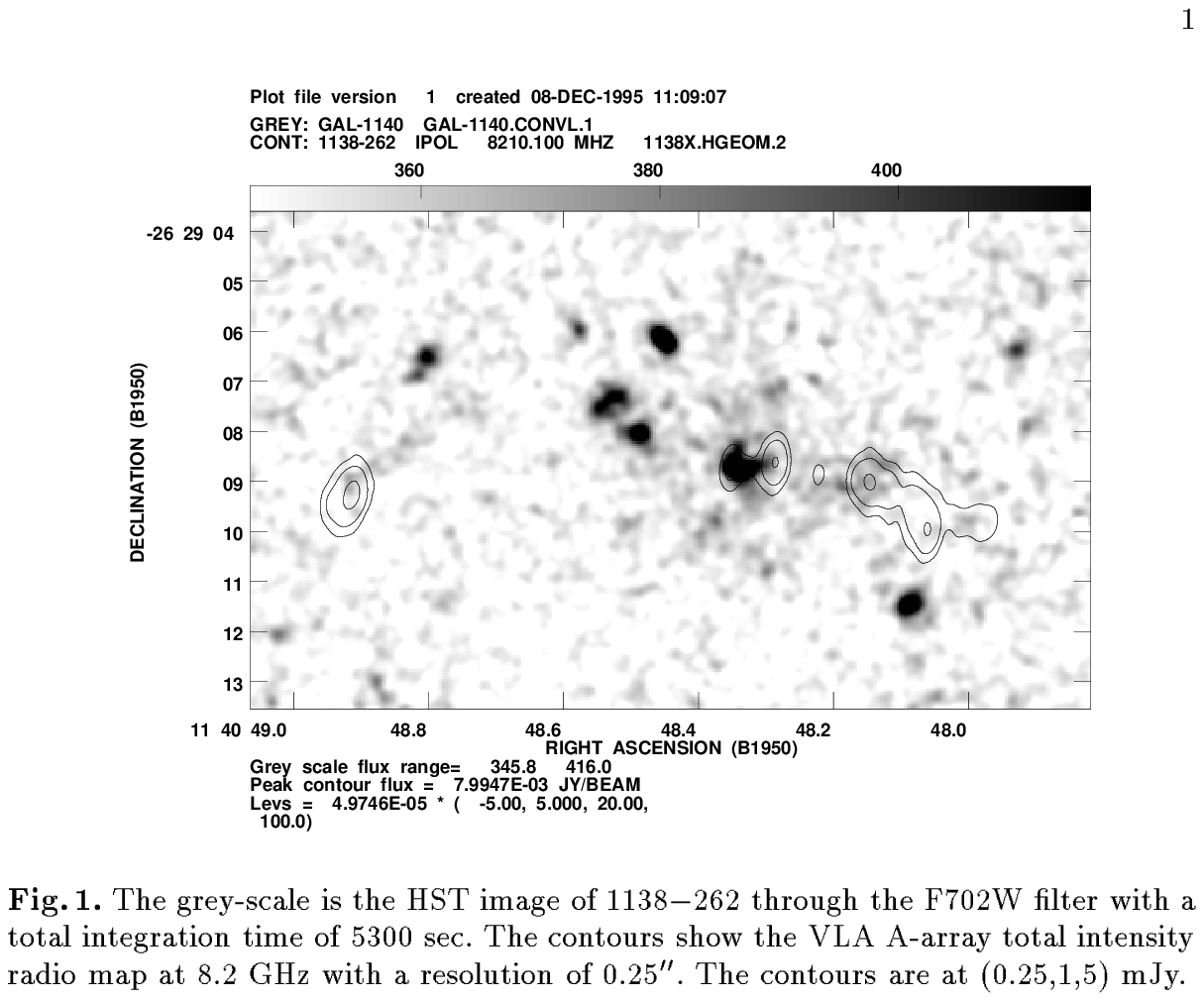,clip=}}
\caption{\label{1138}  
The grey-scale is the HST image of  1138$-$262 through the  F702W filter 
with a total integration time of 5300 sec.
The contours show the VLA A-array total intensity radio map 
at 8.2 GHz  with a resolution of $0.25''$. 
The contours are at (0.25,1,5) mJy.}
\end{figure}

The high resolution imaging capabilities of the HST provide excellent
opportunities for studying the nature of the interaction of the jet
with the host galaxies.  At present we are carrying out an imaging
survey on a selected sample of $z>2$ radio galaxies.  Although the
alignment effect is clearly present at the kpc scale, the diversity
of structures is enormous. Some of the galaxies have a simple cigar
shaped morphologies aligned between the radio lobes (e.g. 0943$-$242,
$z=2.9$, Fig.
\ref{0943}), while others are very complex, showing a number of knots
of which some are connected with the radio jet (e.g. 1138$-$262, $z=2.15$,
Fig. \ref{1138}).

At intermediate redshifts we have studied the optical morphologies as
observed by HST of a complete sample of 3CR radio galaxies ($1 \lta z
\lta 1.3$) and found that they are highly dependent upon their radio
properties \nocite{bes96a} (Best et al. 1996). There is a clear
evolution of the optical structures as the size of the radio source
increases: small radio sources consist of many bright knots, tightly
aligned along the radio axis, whilst more extended sources contain
fewer (generally no more than two) bright components and display more
diffuse emission. 

The morphologies of the intermediate and high-$z$ sources can be
explained as a combination of the three explanations as mentioned
above.  On the basis of the morphologies at the kpc scales most of the
alternative models explaining the alignment effect seem no longer
tenable including (i) inverse Compton scattering of CMB photons (Daly
1992), \nocite{dal92} (ii) enhancement of radio luminosity by
interaction of the jet with an anisotropic parent galaxy (Eales 1992),
\nocite{eal92} (iii) alignment of the angular momentum of the nuclear
black hole with an anisotropic protogalactic distribution (West 1994)
\nocite{wes94} and (iv) gravitational lensing (Le F\`evre et
al. 1987). \nocite{fev87}

If it is indeed correct that the UV/optical continuum can be explained
as due to a combination of scattered, stellar and nebular continuum
light, then the next step is to properly disentangle these three
mechanisms.  Techniques for this include polarisation, colour
information over a large range of wavelengths and narrow band filtering
to separate gas from continuum emission. The superior sensitivity of
the VLT will be particularly important for such photon limited
studies. 

\section{High redshift clusters} 

Detection of clusters and groups of galaxies at high-redshift is important
both for constraining cosmological models and for providing unique
laboratories for studying a diverse range young and forming galaxies.

High-redshift radio sources are important targets for establishing to
what extent clustering exists in the early Universe. At low redshifts
($z \sim 0.1$ to 0.5), luminous steep--spectrum radio sources have long
been 
known to be excellent indicators of galaxy clustering (e.g. Miley
1980). \nocite{mil80b} It has also been shown that high luminosity
radio sources associated with quasars and radio galaxies in the range
$0.5 < z < 1$ are located in rich clusters (e.g. 
Hill and Lilly 1991). \nocite{hil91} 

More recent work clearly indicates that at least some of the radio
galaxies at $z \sim 1$ are contained in clusters.  The most likely
explanation of the X-ray emission detected from  some of the $z\sim 1$
3C radio galaxies (e.g. 3C356, $z=1.079$) Crawford and Fabian 1993) is
that it originates in a hot halo of X-ray gas associated with a
cluster.  At this conference Dickinson presented further evidence from
Keck spectroscopy and ROSAT X-ray imaging that 3C324 ($z=1.206$) is
located in a cluster (see also Dickinson et al. 1995).  \nocite{dic95b} 

At $z>2$ the existence of clusters around HZRGs has not been
established.  There are important observational indications, however, 
that they might be in clusters, including (i) strong Faraday
polarisation and rotation of the radio emission of some of the HZRGs
indicating dense halos of hot electrons (e.g. Carilli et al. 1996),
\nocite{car96a} (ii) an excess of companion galaxies detected along
the axes of the radio sources (R\"ottgering et al.  1996b),
\nocite{rot96d} (iii) deep K-band imaging showing several red
companion galaxies possibly at the same redshift as 4C41.17 ($z=3.8$)
(Graham et al. 1994) \nocite{gra94} and (iv) potential companion
galaxies around 4C41.17 discovered through imaging below the Lyman
limit (Lacy and Rawlings 1996).  \nocite{lac96b}

The deep imaging work that can be carried out (and is being carried out)
with existing 4-m
telescopes provides  good candidates for cluster galaxies around HZRGs.
To establish the existence of a cluster, the  redshifts of these  candidate
cluster galaxies have to be measured.  This will be an important task
for the VLT.

\section{1138$-$262 at $z=2.15$: a young and forming galaxy at the center of cluster?} 

Perhaps the ultimate aim of studying distant radio galaxies is to
investigate the process of galaxy formation.  A number of competing
scenarios of galaxy formation have been proposed.  Here we mention
three classes of models which are currently in vogue and which may be
relevant for such studies.  First there are cooling flow models in
which the galaxy forms during a massive cooling flow. Secondly there
are hierarchical models in which dwarf galaxies merge to form the
large radio galaxy.  Thirdly there are models in which a gas reservoir
builds up and undergoes a massive star burst.

\begin{figure}
\centerline{
\psfig{figure=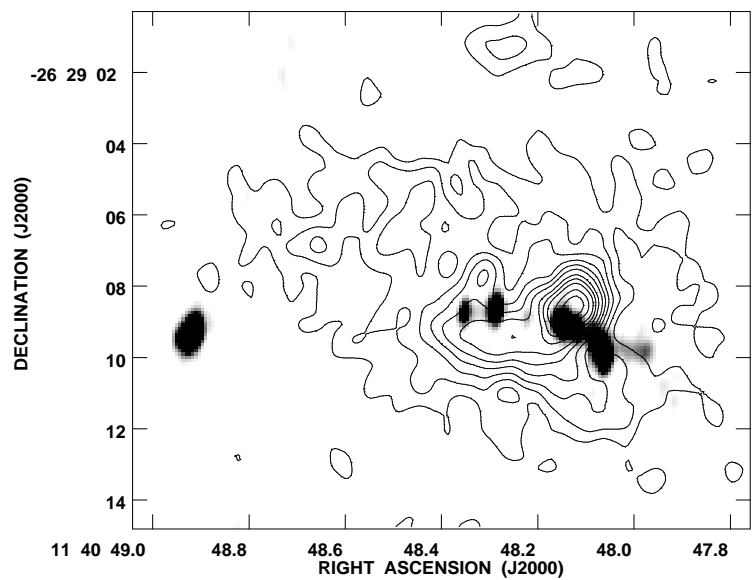,width=10cm}} 
\caption{\label{1138ntt}  The contours show an NTT narrow band image of the
Ly$\alpha$ emission from 1138$-$262. The grey-scale represents 
the VLA A-array total intensity radio map 
at 8.2 GHz  with a resolution of $0.25''$.  } 
\end{figure}

This paper is clearly too short to  discuss in detail these three 
different scenarios. Instead what we would like to 
do here is briefly discuss the remarkable galaxy 
1138$-$262 ($z=2.15$) in the light of these three scenarios. 

A VLA imaging survey of 34 radio galaxies known to be at
$z>2$ has recently been carried out (Carilli et al. 1996)
\nocite{car96a} The  radio galaxy PKS 1138$-$262 ($z=2.15$, see Figs.
\ref{1138} and \ref{1138ntt}) was one of the
most extreme objects in the sample having (i) the highest rotation 
measure (RM)  and the
largest gradient of RM, (ii) the most distorted radio morphology,  and 
(iii) a spectral index that steepens towards the outer knots.  Also
the optical properties are very peculiar with (i) a very clumpy
continuum emission, as seen by HST, and (ii) a distribution of
Ly$\alpha$ emission gas that does not follow the optical continuum
(Pentericci et al. 1996). \nocite{pen96}

Drawing the analogy between low redshift radio galaxies and 1138$-$262
we suggest that this source  is  at the centre of an extreme
cooling-flow, as high as 1000 M$_{\odot} $ yr$^{-1}$. If this cooling 
flow could be maintained long enough then it would indeed 
provide sufficient material from which a large galaxy could be
assembled. The extreme clumpiness of this galaxy suggest 
that that we are witnessing the assembling of this 
galaxy. Once the subunits have merged, 1138$-$262 will be a 
cD type  galaxy at the centre of the cluster. 

The suggestion that 1138$-$262 is a forming galaxy is possibly premature. 
It is clear, however,  that detailed studies with  the 
VLT of objects like this will allow the hypothesis that 
such objects are  protogalaxies to be tested. 
Of particular interest will be a  detailed 
determination of the SED of all the individual clumps of these 
galaxies. What is their stellar content like? 
Are they indeed SF regions that mainly contain 
young stars, or are they fairly old dwarf galaxies 
that happen to be in a cluster around 1138$-$262? 
Such measurements are impossible with current 4-m class telescopes
and should be carried out with the next generation of optical telescopes.

\section{Strategy} 

The VLT
will be equipped with a broad range of instrumentation, the most
important of which are mentioned in Table \ref{instr}.  This table also 
indicates the major emission components of HZRGs 
that can  studied with each
instrument. 

\begin{table} 
\centerline{\psfig{figure=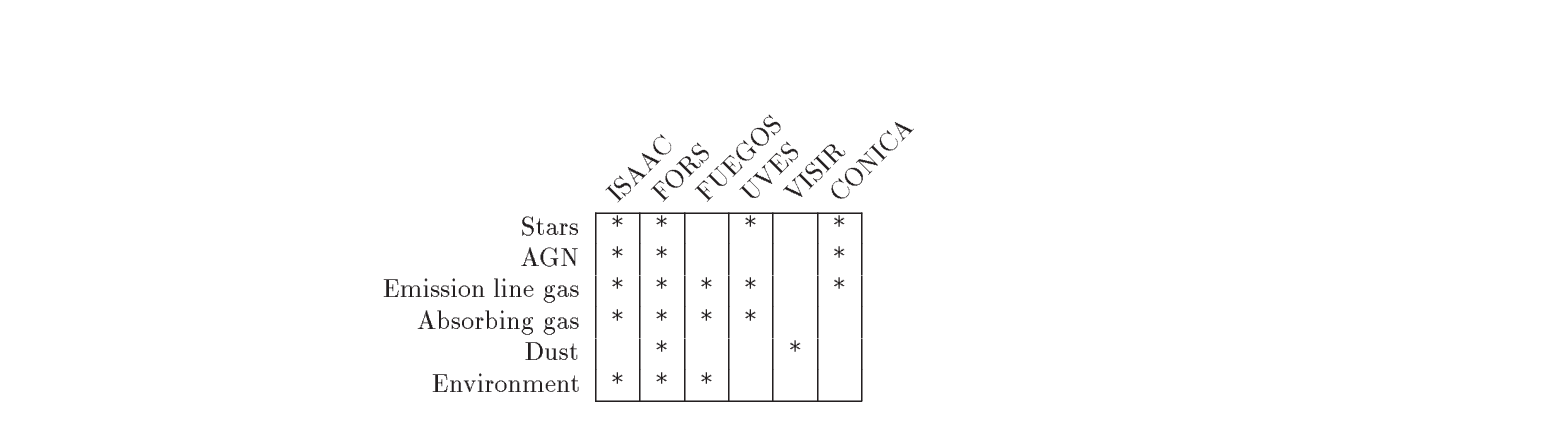} }
\caption{\label{instr} An overview 
of the most important instruments  that the 
VLT will be equipped with, together 
with the components that can be studied with 
these instruments.  } 
\end{table} 

How can we make optimum use of the VLT for these studies?  
A first concern is  the size and quality of samples of distant radio galaxies
that are currently available. 
Preparatory work  should go into a number of  
projects to obtain: (i) significant numbers of $z>4$ radio galaxies,
(ii) complete samples of radio sources that are fully identified and
have redshifts, (iii) samples of milli/$\mu$ Jansky sources and (iv)
samples of radio galaxies near bright stars (for adaptive
optics/VLTI).

A final word about instrumentation: we believe that 
with its first set of instrumentation, the VLT 
is well equipped to carry out studies of the sort that we have mention here. 
We have a slight concern about the number of narrow and 
intermediate band filters that 
will be available. ESO has always had a 
good filter set,  and we hope that 
it finds ways to provide 
an adequate set of filters for the VLT instrumentation. 
Finally, we believe that for studying the 
environment of distant radio galaxies 
a wide field  infrared imaging capability 
is essential. We therefore hope that the present plans 
to build such an instrument will be pursued further.

\bigskip\noindent {\it Acknowledgements.}  We would like to thank our
collaborators, Malcolm Bremer, Wil van Breugel, Chris Carilli, Arjun
Dey, Dick Hunstead, Jaron Kurk, Laura Pentericci, Pat McCarthy, Rob
van Ojik, Hy Spinrad and Paul van der Werf.

%\bibliography{/home/reusel/rottgering/texinputs/huub}
\bibliographystyle{astron}

\end{document}